\documentstyle{article}
\begin{document}
\newcommand{\beq}{\begin{equation}}
\newcommand{\eeq}{\end{equation}}
\newcommand{\beqn}{\begin{eqnarray}}
\newcommand{\eeqn}{\end{eqnarray}}
\newcommand{\dpf}{\displaystyle\frac}
\newcommand{\no}{\nonumber}
\newcommand{\ep}{\epsilon}
\begin{center}
{\Large Entropy, topology of two-dimensional extreme black holes }
\end{center}
\vspace{1ex}
\centerline{\large Bin
Wang$^{a,b,d,}$\footnote[1]{e-mail:binwang@srcap.stc.sh.cn},
\ Ru-Keng Su$^{b,c,}$\footnote[2]{e-mail:rksu@fudan.ac.cn}}
\begin{center}
{$^a$ High Energy Section, International Center for Theoretical Physics,
P.O.Box 586, 34100 Trieste, Italy\\ 
$^b$ China Center of Advanced Science and Technology (World Laboratory),
P.O.Box 8730, Beijing 100080, P.R.China\\
$^c$ Department of Physics, Fudan University, Shanghai 200433,
P.R.China\\
$^d$ Department of Physics, Shanghai Normal University, Shanghai 200234,
P.R.China
}
\end{center}
\vspace{6ex}
\begin{abstract}
Through direct thermodynamic calculations we have shown that different 
classical entropies of
two-dimensional extreme black holes appear
due to two different treatments, namely Hawking's treatment and
Zaslavskii's treatment. Geometrical and topological properties
corresponding to these different treatments are investigated.
Quantum entropies of the scalar fields on the backgrounds of these black
holes
concerning different treatments are also exhibited. Different 
results of entropy and geometry lead us to argue that there are
two kinds of extreme black holes in the nature. Explanation of black
hole phase transition has also been given from the quantum point of
view.

\end{abstract}
\vspace{6ex}
\hspace*{0mm} PACS number(s): 04.70.Dy, 04.20.Gz, 04.62.+v
\newpage
\section{Introduction}

The traditional Bekenstein-Hawking entropy of black hole, which is known to be
proportional to the area $A$ of the horizon, is believed to be appropriate to all
kinds of black holes, including the extreme black hole(EBH). However, recently,
based upon the study of topological properties, it has been argued that the Bekenstein-Hawking
formula of the entropy is not valid for the EBH[1,2]. The entropy of four-dimensional(4D)
extreme Reissner-Nordstrom(RN) black hole is zero regardless of its nonvanishing horizon area.
Further study of the topology [2-4] shew that the Euler characteristic
of such kind of EBH is zero, profoundly different from that of the nonextreme black hole (NEBH).
From the relationship between the topology and the entropy obtained in ref.[4], we see
that this extreme topology naturally results in  the zero entropy.

But these results meet some challenges.  By means of the recaculation of the proper distance
between the horizon and any fixed point, Zaslavskii[5] argued that a 4D RN black hole
in a finite size cavity can approach the extreme state as closely as one likes
but its entropy as well as its temperature on the cavity are not zero. Entropy
is still proportional to the area. Zaslavskii's result has also been supported by
string thoerists got by counting string states [6]. The geometry of 
EBH obtained in the approach of Zaslavskii has been studied in [7,8] and
it
was claimed that its topology is of the nonextreme sector. Meanwhile the string
theorists' results were interpreted by summing over the topology [9],
however this viewpoint has been refuted by Zaslavskii [10,11]. 
These different results indicate that there is a clash
for the understanding of the topology as well as the intrinsic thermodynamical
properties of EBHs. Comparing [1,2] and [5,7,8], this clash seems come
from two
different treatments: one refers to Hawking's treatment by starting with
the original EBH [1,2]
and the other Zaslavskii's treatment by first taking the boundary limit
and then the extreme
limit to get the EBH from nonextreme counterpart [5,7,8]. Recently we
have studied the geometry and intrinsic thermodynamics of extreme Kerr
black hole by using these two treatments and found that these different
treatments approach to two different topological objects and lead to
drastically different intrinsic thermodynamical properties [12]. Of
course, all these results obtained are limited in the classical treatment
of 4D black holes.

The motivation of the present paper is to extend 4D classical studies
to two-dimensional (2D) black holes. We hope that the
mathematical simplicity in 2D black holes can help us to understand the
problem clearer and
deeper. The results on 2D charged dilaton black hole topology and
thermodynamics were only announced and briefly summarized in [13,14]. To 
make the study general, we will study two kinds of 2D black holes,  the 2D
charged dilaton 
black hole [15,16] as well as the 2D Lowe-Strominger black hole [17], by
using two treatments mentioned above in detail.
We will prove that these two treatments result in two different thermal
results:
Bekenstein-Hawking entropy and zero entropy for EBHs. Besides we will
investigate the geometry and topology of 2D EBHs in detail and
directly relate two treatments to topological properties of EBHs.
We will clearly exhibit Euler characteristic values for 2D EBHs,
especially for EBHs obtained from the nonextreme counterpart by
Zaslavskii's treatment.
Different Euler characteristics
are directly derived from different treatments, rather than by introducing
some other conditions, such as the inner boundary condition as done in 4D
cases [3].

The other objective of the present paper is to study this problem quantum
mechanically. As early pointed out by t'Hooft[18], the fields propagating
in the region just outside the horizon give the main contribution to the
black hole entropy. Many methods, for example,
the brick wall model[18,19], Pauli-Villars regular theory[20] etc., have
been suggested to study the quantum effects of entropy under WKB approximation
or one-loop approximation. Suppose the black hole is enveloped by a scalar
field, and the whole system, the hole and the
scalar field, are filling in a cavity. Adopt the viewpoint that the entropy
arises from entanglement[21-23], it is of interest to study the quantum
effects
of these two different treatments on the entropy of the scalar field
on the EBHs' backgrounds under WKB approximation, in particular, to
investigate whether these two different treatments will offer two different
values of entropy. In Sec.IV we will prove that the entropy of the scalar field
depends on two different treatments as well. 
Some physical understanding concerning these results will also be given.

The organization of the paper is as the following: In Sec.II, the classical
entropy of two kinds of 2D EBHs are derived by using two different
treatments.
And in Sec.III, the geometry and topology of these EBHs are investigated. The
Euler characteristics are clearly exhibited. Sec.IV is devoted to the discussion of the
entropy of the scalar field on the EBH background. The conclusions and discussions
will be presented in the last section.

\section{Classical entropy}

We first study the 2D charged dilaton black hole(CDBH)[15,16]. The action
is
\beq                  
I=-\int_M \sqrt{g}e^{-2\phi}[R+4(\bigtriangledown\phi)^2+\lambda^2-\dpf{1}{2}
  F^2]-2\int_{\partial M}e^{-2\phi}K
\eeq
which has a black hole solution with the metric
\beqn         
{\rm d}s^2=-g(r){\rm d}t^2+g^{-1}(r){\rm d}r^2\\
g(r)=1-2me^{-\lambda r}+q^2e^{-2\lambda r}      \\
e^{-2\phi}=e^{-2\phi_0}e^{\lambda r},\ A_0=\sqrt{2}qe^{-\lambda r}
\eeqn
where $m$ and $q$ are the mass and electric charge of the black hole respectively.
The horizons are located at $r_{\pm}=(1/\lambda)\ln(m\pm \sqrt{m^2-q^2})$.

Using the finite-space formulation of black hole thermodynamics, employing the
grand canonical emsemble and putting the black hole into a cavity as
usual[5,24,25],
we calculate the free energy and entropy of the CDBH. To simplify our calculations, we
introduce a coordinate transformation
\beq    
r=\dpf{1}{\lambda}\ln[m+\dpf{1}{2}e^{\lambda(\rho+\rho_0^*)}+\dpf{m^2-q^2}{2}
    e^{-\lambda(\rho+\rho_0^*)}]
\eeq
where $\rho_0^*$ is an integral constant, and rewrite Eq.(2) to a particular gauge
\beq     
{\rm d}s^2=-g_{00}(\rho){\rm d}t^2+{\rm d}\rho^2
\eeq
The Euclidean action takes the form
\beq         
I=-\int_{\partial M} \sqrt{\dpf{1}{g_{11}}}e^{-2\phi}(\dpf{1}{2}\dpf{\partial _1 g_{00}}
   {g_{00}}-2\partial_1 \phi)
\eeq
The dilaton charge is found to be
\beqn          
D=e^{-2\phi_0}(m+\dpf{1}{2}e^x+\dpf{m^2-q^2}{2}e^{-x})\\
x=\lambda(\rho+\rho_0^*)
\eeqn
The free energy, $F=I/{\beta}$, where $\beta$ is the proper periodicity of
Euclideanized time at a fixed value of the special coordinate and has the form
$\beta=1/T_w=\sqrt{g_{00}}/T_c$, $T_c$ is the inverse periodicity of the Euclidean time
at the horizon
\beq         
T_c=\dpf{\lambda\sqrt{m^2-q^2}}{2\pi(m+\sqrt{m^2-q^2})}
\eeq
Using the formula of entropy
\beq        
S=-(\partial F/\partial T_w)_D
\eeq
 we obtain
\beq             
S=\dpf{2\pi e^{-2\phi}[m+\dpf{e^x}{2}+\dpf{(m^2-q^2)e^{-x}}{2}][1+(m^2-q^2)e^{-2x}]
   \sqrt{m^2-q^2}(m+\sqrt{m^2-q^2})}{(m^2-q^2)+m[\dpf{e^x}{2}+\dpf{(m^2-q^2)e^{-x}}{2}]}
\eeq
Taking the boundary limit $x\rightarrow x_+=\lambda(\rho_+ +\rho_0^*)=\ln\sqrt{m^2-q^2}$
in Eq.(12) to get the entropy of the hole, we find
\beq         
S=4\pi e^{-2\phi_0}(m+\sqrt{m^2-q^2})
\eeq
This is just the result given by Nappi and Pasquinucci[15] for the
non-extreme CDBH,
which confirms that our treatment above is right.

   We are now in a position to entend the above calculations to EBH. We are facing
two limits, namely, the boundary limit $x\rightarrow x_+$ and the extreme limit
$q\rightarrow m$. We can take the limits in different orders: (A) by first taking
   the boundary limit $x\rightarrow x_+$, and then the extreme limit
$q\rightarrow m$ as the treatment adopted in [5,7,8];
   and (B) by first taking the extreme limit $q\rightarrow m$ and then the boundary
   limit $x\rightarrow x_+$, which corresponds to the treatment of Hawking
et al. [1,2] by starting with the original EBH.
   To do our limits procedures mathematically,
   we may take $x=x_+ +\ep, \ep\rightarrow 0^+$ and $m=q+\eta, \eta\rightarrow 0^+$,
   where $\ep$ and $\eta$ are infinitesimal quantities with different orders of magnitude,
   and substitute them into Eq.(12). It can easily be shown that in treatment (A)
   \beq          
   S_{CL}(A)=4\pi me^{-2\phi_0}
   \eeq
   which is just the Bekenstein-Hawking entropy. However, in treatment (B),
   \beq            
   S_{CL}(B)=0
   \eeq
   which is just the result given by refs.[1,2].

These peculiar results can also be found in  
2D Lowe-Strominger black hole. This 2D black hole is obtained in [17] by
instroducing
gauge fields through the dimensional compactification of three-dimensional string
effective action. The 2D action in this case has the form
\beq               
I=-\int_{M}\sqrt{g}e^{-2\phi}[R+2\lambda^2-\dpf{1}{4}e^{-4\phi}F^2]-2\int_{\partial M}e^{-2\phi}K
\eeq
where $\phi$ is a scalar field coming from the compactification and plays the role of
dilaton for the 2D action. This action possesses the black hole solution
\beqn                
{\rm d}s^2 & = & -(\lambda^2r^2-m+\dpf{J^2}{4r^2}){\rm d}t^2+\dpf{1}{\lambda^2r^2-m+\dpf{J^2}{4r^2}}{\rm d}r^2 \\
       A_0 & = & -\dpf{J}{2r^2} \\
e^{-2\phi} & = & r
\eeqn
The parameter $J$ in this solution gives ``charge" to this black hole[25].
The horizons of
this black hole locate at
\beq          
r_{\pm}=\dpf{1}{\lambda}\{\dpf{m}{2}[1\pm(1-(\dpf{\lambda J}{m})^2)^{1/2}]\}^{1/2}
\eeq
where $r_+$ is the event horizon and $r_-$ the inner cauchy horizon. In the extreme
limit $\lambda J\rightarrow m$, $r_+$ and $r_-$ degenerate.

Using the finite space formulation of black hole thermodynamics, employing the grand
canonical ensemble and putting the hole into a cavity as usual[5,24], we
calculate
the free energy and the entropy of the hole. As done in 2D CDBH, we
introduce a coordinate transformation to simplify the calculation
\beq           
r^2=\dpf{1}{2\lambda^2}(m+\dpf{1}{2}e^{2\lambda(\rho+\rho_0)}+\dpf{m^2-\lambda^2 J^2}{2}e^{-2\lambda(\rho+\rho_0)})
\eeq
where $\rho_0$ is an integral constant, and rewrite Eq(17) to a particular gauge Eq(6).
After transformation, the event horizon locates at
\beq           
\rho_+=\dpf{1}{2\lambda}\ln\sqrt{m^2-\lambda^2 J^2}-\rho_0
\eeq
The Euclidean action can be evaluated as
\beq             
I=-\int_{\partial M}[n^aF_{ab}A^be^{-6\phi}+2Ke^{-2\phi}]
\eeq
The free energy can be obtained by the evaluation of (23). Employing the constant
shift in the gauge potential $A_a\rightarrow A_a+constant$, in the equations of motion to avoid
divergence in the gauge potential at the horizon as done in [25,26], we
have the free energy
\beq        
F=-2\lambda D\dpf{e^{2x}+(m^2-\lambda^2 J^2)e^{-2x}+2\sqrt{m^2-\lambda^2 J^2}}{e^{2x}-(m^2-\lambda^2 J^2)e^{-2x}}
\eeq
where $x=\lambda(\rho+\rho_0)$, $D$ is the dilaton charge given by
\beq            
D=[\dpf{1}{2\lambda^2}(m+\dpf{1}{2}e^{2x}+\dpf{m^2-\lambda^2 J^2}{2}e^{-2x})]^{1/2}
\eeq
Using Eq.(11), where $T_w$ is the inverse periodicity of the Euclideanized time at a fixed value of the special coordinate.
$T_c$ is the inverse periodicity at the horizon, reads
\beq             
T_c=\dpf{\sqrt{2}\lambda\sqrt{m^2-\lambda^2 J^2}}{2\pi(m+\sqrt{m^2-\lambda^2 J^2})^{1/2}}
\eeq
We find the entropy as
\beqn             
S & = & \dpf{4\pi\sqrt{m^2-\lambda^2 
J^2}(m+\dpf{e^{2x}}{2}+\dpf{m^2-\lambda^2
 J^2}{2}e^{-2x})e^{-2x}(\dpf{e^{2x}}{2}+\dpf{m^2-\lambda^2
J^2}{2}e^{-2x}+\sqrt{m^2-\lambda^2 J^2})}
  {\sqrt{2}\lambda[(m+\dpf{e^{2x}}{2}+\dpf{m^2-\lambda^2 
J^2}{2}e^{-2x})^2-\lambda^2 J^2]} \\ \no
&   & \times (m+\sqrt{m^2-\lambda^2 J^2})^{1/2}
\eeqn
Taking the boundary limit $$x\rightarrow x_+=\lambda(\rho_+ +\rho_0)=\dpf{1}{2}\ln\sqrt{m^2-\lambda^2 J^2}$$
in Eq(27), one has
\beq              
S=\dpf{4\pi}{\sqrt{2}\lambda}(m+\sqrt{m^2-\lambda^2 J^2})^{1/2}
\eeq
This is just the result given in [25] for the NEBH.

As in the 2D CDBH we are facing two limits to extend the above calculation to EBH,
namely, the boundary limit $x\rightarrow x_+$ and the extreme limit
$\lambda J\rightarrow m$. And again there are two treatments:
(A) Zaslavskii's treatment by first taking the boundary limit
$x\rightarrow x_+$, and then the extreme limit $\lambda J\rightarrow m$; and (B)
Hawking et al.'s treatment by first taking the extreme limit $\lambda
J\rightarrow m$ and then the boundary limit $x\rightarrow x_+$.
It is easy to find
for treatment (A)
\beq           
S_{CL}(A)=\dpf{2\sqrt{2}\pi}{\lambda}m^{1/2}
\eeq
This is just the Bekenstein-Hawking entropy. It depends on the mass $m$ only. 

However, in treatment (B), we find
\beq              
S_{CL}(B)=0
\eeq
This is just the result given by refs[1,2].

Therefore, through direct thermodynamical
calculations for two kinds of extreme 2D black holes, we come to a conclusion that
the different results of entropy in fact
come from two different treatments.

Now we are facing a puzzle. In statistical physics and thermodynamics, entropy
is a function of state only, and does not depend on the history or the process
of how the system arrives at the equilibrium state as well as the different treatment of
mathematics. But now we use two different treatments for orders of limits to arrive at
the final state, namely, the EBH state statisfying the same extreme
condition, we get two different values of the entropy.
The puzzle is similar to that in 4D RN cases and need further discussions.

\section{Geometry and topology}

 In this section, we
study the relation between the geometrical properties and two different
treatments
of taking different limits in detail.

Consider the Euclidean metric of 2D CDBH
\beq     
{\rm d}s^2=g(r){\rm d}\tau^2+g(r)^{-1}{\rm d}r^2
\eeq
where $g(r)$ has the form of Eq.(3). Taking the new variable $\tau_1=2\pi T_c\tau$
where $0\leq \tau_1\leq 2\pi$, then Eq(31) becomes
\beq        
{\rm d}s^2=(\beta/2\pi)^2{\rm d}\tau_1^2+{\rm d}l^2
\eeq
$\beta$ is the inverse local temperature and $l$ is the proper distance. The equilibrium
condition of the spacetime reads
\beq                 
\beta=\beta_0[g(r_B)]^{1/2}, 1/{\beta_0}=T_c=g'(r_+)/{4\pi}
\eeq
Let us choose the coordinate according to
\beq          
r-r_+=4\pi T_c b^{-1}\sinh^2(x/2), b=g'(r_+)/2
\eeq

For the treatment (A), taking the limit $r_+\rightarrow r_B$ first, where the hole tends to occupy the entire cavity, the
region $r_+\leq r\leq r_B$ shrinks and we can expand $g(r)$ in the power
of series
$g(r)=4\pi T_c(r-r_+)+b(r-r_+)^2+\cdots$ near $r=r_+$. After substitution
into
Eqs.(32,33), and take the extreme limit $r_+=r_-=r_B$, $b=\lambda^2$, Eq(31) can be expressed as
\beq         
{\rm d}s^2(A)=\lambda^{-2}(\sinh^2x{\rm d}\tau_1^2+{\rm d}x^2)
\eeq
This is just the 2D counterpart of the Bertotti-Robinson(BR) spacetime
[27].
However, for the treatment (B), we start from the original extreme 2D CDBH, $g(r)=(1-me^{-\lambda r})^2$.
Introducing the variable $r-r_+=r_B\rho^{-1}$ and expand the metric coefficients
near $r=r_+$, we obtain
\beq         
{\rm d}s^2(B)=\rho^{-2}(\lambda^2r_B^2{\rm d}\tau^2+{\rm d}\rho^2/\lambda^2)
\eeq

Now we are in a position to discuss the properties of Eqs(35,36). The horizons of the EBH
got in the treatment (A) is determined by
\beq          
g=1/4g'(r_+)b^{-1}\sinh^2x=0
\eeq
So the horizon locates at finite $x$, say $x=0$. The proper distance between the
horizon and any other point is finite. However for the original extreme 2D CDBH (36), the horizon is detected by
\beq                 
g=\lambda^2r_B^2\rho^{-2}=0
\eeq
therefore, the horizon is at $\rho=\infty$. The distance between the horizon and any other $\rho<\infty$
is infinite. It is this difference here that gives rise to the qualitatively
different topological features of the EBHs.

To exhibit these different topological features, we calculate the Euler characteristic of these two EBHs directly.
The formula for the calculation of the Euler characteristic in 2D cases
is[28]
\beq                
\chi=\dpf{1}{2\pi}\int R_{1212}e^1\wedge e^2
\eeq
For the nonextreme 2D CDBH, adopting its metric, 
and substracting the asymptotically flat space's influence [3,28], we
arrive at
\beq    
\chi=-\dpf{1}{2\pi}\beta_0[-m\lambda e^{-\lambda r}+q^2\lambda e^{-2\lambda r}]_{r_+}=1
\eeq
This result is in accordance with that of the multi-black-holes obtained in [3].

It is easy to extend the calculation of $\chi$ to the cases of EBHs. For the EBH developed
from the treatment (A), the Euler characteristic can be
directly got by taking $r_+=r_B$ first and $m=q$ afterwards,
\beqn         
\chi(A) & = & -\dpf{1}{2\pi}\beta_0[-m\lambda e^{-\lambda r}+q^2\lambda e^{-2\lambda r}]_{r_+=r_B}\vert_{extr} \no \\
     & = & \dpf{2\pi(m+\sqrt{m^2-q^2})\lambda\sqrt{m^2-q^2}}{2\pi\lambda\sqrt{m^2-q^2}
     (m+\sqrt{m^2-q^2})}\vert_{extr}=1
\eeqn
The same as that of the NEBH.
However for the original extreme 2D CDBH, using the limit procedure (B) and from Eq.(40), we have
\beq         
\chi(B)=-\dpf{1}{2\pi}\beta_0[-m\lambda e^{-\lambda r}+m^2\lambda e^{-2\lambda r}]_{r_+}
\eeq
The horizon for the original EBH is $r_+=\dpf{1}{\lambda}\ln m$, so
\beq           
\chi(B)=0
\eeq
It is quite different from that of the NEBH.

In order to obtain the property in gerneral, we preceed our discussion to
2D Lowe-Strominger black hole. The Euclidean metric has the same form as Eq.(31),
but now
\beq            
g(r)=-m+\lambda^2r^2+\dpf{J^2}{4r^2}=\dpf{\lambda^2}{r^2}(r^2-r_+^2)(r^2-r_-^2)
\eeq
Using the same treatment (A) as that of 2D CDBH,
namely, taking the boundary condition first and then the extreme limit, we find that the
metric can be expressed as
\beq        
{\rm d}s^2(A)=(4\lambda^2)^{-1}({\rm d}\tau_1^2\sinh^2x+{\rm d}x^2)
\eeq
It is a 2D counterpart of BR spacetime again. While starting from the original EBH,
$g(r)=\dpf{\lambda^2}{r^2}(r^2-r_+^2)^2$, as the original extreme 2D CDBH, the metric
can be written as
\beq              
{\rm d}s^2(B)=\rho^{-2}(4\lambda^2r_B^2{\rm d}\tau^2+\dpf{1}{4\lambda^2}{\rm d}\rho^2)
\eeq
The location of the horizon can be found directly for these two expressions of metric.
For Eq.(45), \\ from $g=\dpf{1}{4}g'(r_+)b^{-1}\sinh^2x=0$, the horizon is at finite $x$,
say $x=0$. While for Eq.(46), the horizon is determined by $g=4\lambda^2r_B\rho^{-2}=0$,
so it is at $\rho=\infty$. The proper distances between a horizon and
any other point are finite and infinite for Eqs.(45,46) respectively.

Applying Eq(39), we can also get the results for the Euler characteristic for these
two kinds of 2D Lowe-Strominger black holes. Substituting the metric in Eq(39), the Euler characteristic for the NEBH reads
\beq            
\chi=-\dpf{\beta_0}{2\pi}[-\lambda^2r+\dpf{J}{4r^3}]_{r_+}=1
\eeq
where we have used Eq(17) and (20). The Euler characteristic for the EBHs are
obvious. For the EBH got in the treatment (A), 
we obtain
\beq          
\chi(A)=-\dpf{\beta_0}{2\pi}[-\lambda^2r+\dpf{J^2}{4r^3}]_{r_+=r_B}\vert_{extr}=1
\eeq
However, in treatment (B), we find
\beq         
\chi(B)=-\dpf{\beta_0}{2\pi\lambda^2}[-\lambda^4r+\dpf{m^2}{4r^3}]_{r_+}
\eeq
Considering $r_+^2=\dpf{m}{2\lambda^2}$ for the original EBH, we finally get
\beq        
\chi(B)=0
\eeq

These results clearly show that the different treatments result in
different geometrical
 and topological properties. The direct relation to the priority of taking
limits,
rather than introducing additional condition[3], makes the outcomes more
concise and explicit. Different topological results obtained here make us
easier to accept
the different classical entropy derived for 2D EBHs. We find that in
addition to 4D cases claimed in [3,4,12], in the 2D cases, the topology
and
the EBHs'
classical entropy are closely related.

\section{Quantum entropy}

     An early suggestion by 't Hooft[18] was that the fields propagating
in the region
      just outside the horizon give the main contribution to the black
hole entropy.
      The entropy of the black hole system arises from
entanglement[21-23]. Many methods, for example, the brick wall
     model[12,13], Pauli-Villars regular theory[20],etc., have been
suggested to calculate
      the quantum effects of entropy in WKB approximation or in the
one-loop approximation.
Let's first study the 2D CDBH and
suppose the CDBH is enveloped by a scalar field, and the whole system, the hole and
     the scalar field, are filling in a cavity. The wave equation of the scalar field is
     \beq           
     \dpf{1}{\sqrt{-g}}\partial _\mu(\sqrt{-g}g^{\mu\nu}\partial _\nu\phi)-M^2\phi=0
     \eeq
     Substituting the metric Eq.(2) into Eq.(51), we find
     \beq          
     E^2(1-2me^{-\lambda r}+q^2e^{-2\lambda r})^{-1}f+\dpf{\partial}{\partial r}
      [(1-2me^{-\lambda r}+q^2e^{-2\lambda r})\dpf{\partial f}{\partial r}]-M^2f=0
      \eeq
      Introducing the brick wall boundary condition[18]
      \beqn
      \phi(x)=0\  {\rm at}\ r=r_+ +\ep\no     \\
      \phi(x)=0\  {\rm at}\ r=L  \no   \\  \no
      \eeqn
      and calculating the wave number $K(r,E)$ and the free energy $F$, we get
\beq                    
K^2(r,E)=(1-2me^{-\lambda r}+q^2e^{-2\lambda r})^{-1}[(1-2me^{-\lambda r}+q^2e^{-2\lambda r})^{-1}E^2-M^2]
\eeq
\beq             
F_{QM}=\dpf{\pi}{6\beta^2 \lambda}[\dpf{1}{2}\ln(R^2-2mR+q^2)+\dpf{m}{2\sqrt{m^2-q^2}}
	\ln\dpf{R-m-\sqrt{m^2-q^2}}{R-m+\sqrt{m^2-q^2}}]
\eeq
where $R=e^{\lambda(r_+ +\ep)}$, and $\ep\rightarrow 0$ is the coordinate cutoff
parameter.
To extend the above discussion to EBH, we are facing two limits $\epsilon\rightarrow 0$ and $q\rightarrow m$ again.
It can be proved that Eq(54) depends on the order of taking these two limits.
We find for treatment (A) which we first take boundary limit
$\ep\rightarrow 0$ and then the extreme limit $q\rightarrow m$
\beq                 
F_{QM}(A)=-\dpf{\pi}{6\beta^2 \lambda}\ln(\dpf{1}{m\lambda \epsilon}+
\dpf{m}{2\sqrt{m^2-q^2}}\ln\dpf{2\sqrt{m^2-q^2}}{\lambda\epsilon(m+\sqrt{m^2-q^2})})
\eeq
We just leave the term $\sqrt{m^2-q^2}$ in the second term of Eq.(55) for
the moment for the following discussions.

But for treatment (B) by first adopt the extreme limit and then the
boundary limit
\beq              
F_{QM}(B)=-\dpf{\pi}{6\beta^2 \lambda}(\dpf{m}{m\lambda \epsilon} 
+\ln\dpf{1}{m\lambda \epsilon})
\eeq
 Similar to the classical case, different expressions for free energy
appear here due to different priority of taking different limits.
Through the entropy formula $S=\beta^2(\partial F/\partial\beta)$, we obtain
\beqn              
S_{QM}(A)=\dpf{\pi}{3\beta\lambda}(\ln\dpf{1}{m\lambda\ep}+\dpf{m} 
{2\sqrt{m^2-q^2}}\ln\dpf{2\sqrt{m^2-q^2}}{\lambda\ep(m+\sqrt{m^2-q^2})})\\
S_{QM}(B)=\dpf{\pi}{3\beta\lambda}(\dpf{m}{m\lambda\ep}+\ln\dpf{1}{m\lambda\ep})
\eeqn
We conclude that the entropy on the black hole background also depends on 
two different treatments.

Now we turn to study 2D Lowe-Strominger model.
We suppose the 2D black hole is enveloped by a scalar field, and
the whole system are filling in a cavity. 
Substituting the metric Eq(17) into Eq(51),
we get the radial equation as
\beq             
E^2(-m+\lambda^2 r^2+\dpf{J^2}{4r^2})^{-1}f+\dpf{\partial}{\partial r}[(-m+\lambda^2 r^2+\dpf{J^2}{4r^2})\dpf{\partial f}{\partial r}]
-M^2f=0
\eeq
The wave number is
\beq        
K^2=(-m+\lambda^2 r^2+\dpf{J^2}{4r^2})^{-1}[(-m+\lambda^2 r^2+\dpf{J^2}{4r^2})^{-1}E^2-M^2]
\eeq
and the semiclassical quantization condition is
\beq               
n\pi=\int^{L}_{r_+ +\ep}{\rm d}r K(r,E)
\eeq
The free energy satisfies
\beqn                  
\beta F &=&\sum_{n}\log(1-e^{-\beta E}) \\ \no
	&=& \int{\rm d}n\log(1-e^{-\beta E}) \\ \no
	&=& -\dpf{\beta}{\pi}\int{\rm d}E(e^{\beta E}-1)^{-1}f(r)
\eeqn
where
\beq           
f(r)=\int^{L}_{r_+ +\ep}{\rm d}r(-m+\lambda^2 r^2+\dpf{J^2}{4r^2})^{-1}\sqrt{E^2-M^2(-m+\lambda^2 r^2+\dpf{J^2}{4r^2})}
\eeq
Expanding to the powers of $M$ in the limit $\ep\rightarrow 0$, the leading contribution can
be obtained as
\beq          
f(r)_{r=r_+ +\ep}=-\dpf{1}{\lambda^2}(\dpf{A}{2r_+}\ln\dpf{\ep}{2r_+ +\ep}+\dpf{B}{2r_-}\ln\dpf{r_+-r_-+\ep}{r_++r_-+\ep})
\eeq
where
\beq               
A=\dpf{r_+^2}{r_+^2-r_-^2}, \  B=-\dpf{r_-^2}{r_+^2-r_-^2}
\eeq
For treatment (A),
\beq               
F(A)=-\dpf{\pi}{6\beta^2}(\dpf{1}{4r_+\lambda^2}\ln\dpf{2r_+}{\ep}+
\dpf{r_-}{2\lambda^2(r_+^2-r_-^2)}\ln\dpf{2r_+(r_+-r_-)}{\ep(r_+-r_-)})
\eeq
As in Eq.(55), we leave $r_+ -r_-$for the moment.
 
For treatment (B), the free
energy is
\beq                  
F(B)=-\dpf{\pi}{6\beta^2}(\dpf{1}{4\lambda^2\ep}+\dpf{1}{4r_+\lambda^2}\ln\dpf{2r_+}{\ep})
\eeq
Using the thermodynamic formula
we find
\beqn           
S_{QM}(A)=\dpf{\pi}{3\beta\lambda^2}(\dpf{1}{4r_+}\ln\dpf{2r_+}{\ep}+\dpf{r_-}
{2(r_+^2-r_-^2)}\ln\dpf{2r_+(r_+-r_-)}{\ep(r_+-r_-)})
\\
S_{QM}(B)=\dpf{\pi}{3\beta\lambda^2}(\dpf{1}{4\ep}+\dpf{1}{4r_+}\ln\dpf{2r_+}{\ep})
\eeqn
respectively.

We find that two different values of entropies of scalar field on the
backgrounds of extreme holes
exhibit again. The results of 2D CDBH and Lowe-Strominger black holes tell us that the entropy of
the scalar field on the EBHs' background depend on the limits procedures as well.

\section{Conclusions and discussions}

Through direct thermodynamic calculations, we have shown that corresponding to
two different treatments of the orders of taking the boundary limit and
black hole extreme limit,
the classical entropies of 2D extreme CDBH and Lowe-Strominger black hole have different values,
namely, zero and Bekenstein-Hawking value. We have shown that the geometrical and topological properties
are also dependent on these two different treatments by direct provement.
Since the extreme conditions are all
satisfied for the discussed EBHs, these profoundly different goemetrical
and topological
properties lead us to an impression that there are two kinds of EBHs which are classified
by different topologies. One is the original EBH as Hawking et al. claimed
and due to peculiar topology of this EBH, this kind of EBH cannot be
formed from their nonextreme counterpart and can only be prepared in the
early universe. While the other EBH obtained by treatment (A) with the
same topology as that of
NEBH can be developed from their NEBH counterpart. 
Entropies and topological properties for EBHs
studied in 4D cases[3,4,12] as well as in 2D cases in our paper suggest
that there are close relation
between the topology and the classical entropy for EBHs. 
Our results
have been supported recently by Pretorius, Vollick and Israel [31].

Using the brick wall model, we have shown that
under WKB approximation the entropy of scalar field on the background of extreme 2D CDBH and
Lowe-Strominger black hole
have two values given by Eqs(57,58) and (68,69), respectively.
These results support our argument from the  quantum point of view that
the backgrounds of EBHs are different due to different
topologies.

Besides the usual ultraviolet divergence $\ep\rightarrow 0$, which has
been found for both 4D and 2D NEBH cases [18-23] and was suggested capable
of being overcome by diferent renormalization methods [29,30], a new
divergent term for Zaslavskii's treatment emerges in Eqs.(57,68) and is
absent in Hawking's treatment. To understand this difference, let's go
over the arguments of Hawking et al. and Zaslavskii again. Owing to
different topological properties, Hawking et al.claimed that their EBH and
its NEBH counterpart are completely different objects and their EBH is the
original EBH, which can only be prepared at the beginning of the universe.
In Hawking's treatment, since we put $m=q $ and $r_+=r_-$ first for 2D
CDBH and Lowe-Strominger
black hole respectively, and then calculate their entropies by using 
usual thermodynamical approachs [24], naturaly their quantum entropy
includes ultraviolet divergence only. But for the EBH obtained by using
Zaslavskii's treatment, it can be developed from the nonextreme
counterpart
by taking the extreme limits, therefore new divergent terms appear
here. Keeping in mind that the entropy of scalar fields in the EBH
background was derived by WKB approximation, the divergent quantum entropy
due to taking extreme limits reflects in fact the divergent quantum
fluctuations of the entanglemnt entropy of the whole system including the
EBH and the scalar field. In statistical physics we know that infinite
fluctuation breaks down the rigious meanings of thermodynamical quantities
and is just the characteristic of the point of phase transition. This
conclusion is in good agreement with the previous studies about the phase
transition of black holes by using Landau-Lifshitz theory [32-34]. So a
phase transition will happen when a NEBH approaches to the EBH with
nonextreme topology obtained by
using Zaslavskii treatment. The new divergent terms appears in the quantum
entropy here gives the quantum understanding of phase transition and
supports previous classical arguments.

B. Wang is greatful to Prof. S. Randjbar-Daemi for inviting him to visit
ICTP (Italy), where part of
this work was done. This work was supported by Shanghai Higher Education
and Shanghai Science and Technology Commission.

\end{document}